\documentstyle[epsfig]{elsart} 
\newbox\mybox 
\newcommand\fverb{\setbox\mybox=\hbox\bgroup\verb} 
\newcommand\fverbdo{\egroup\medskip\noindent\fbox{\unhbox\mybox}\ } 
\newcommand\fverbit{\egroup\item[\fbox{\unhbox\mybox}]} 
\newcommand\init[1]{\setbox\mybox=\hbox{{\beeg #1}~}% 
                   \noindent\global\hangindent=\wd\mybox\global\hangafter-2% 
                   \sc\smash{\llap {\lower 13.2pt \box\mybox}}} 
\def\v1{\vspace{1cm}} 
\def\be{\begin{equation}} 
\def\ee{\end{equation}} 
\def\bc{\begin{center}} 
\def\ec{\end{center}} 
\def\vh{\varphi}

\newcommand{\bea}{\begin{eqnarray}} 
\newcommand{\eea}{\end{eqnarray}}

\begin{document} 
\begin{frontmatter} 
\title{Description of Supernova Data in Conformal Cosmology without 
Cosmological Constant} 
\author[ur,dub]{Danilo Behnke,} 
\author[ur,dub]{David B. Blaschke,} 
\author[dub]{Victor N. Pervushin,} 
\author[dub]{Denis Proskurin} 
\address[ur]{Fachbereich Physik, Universit\"at Rostock, D-18051 Rostock, 
Germany} 
\address[dub]{Bogoliubov Laboratory for Theoretical Physics, JINR, 
141980 Dubna, 
Russia} 
\begin{abstract} 
We consider cosmological consequences of a conformal - invariant 
formulation of Einstein's General Relativity where instead of the scale 
factor of the spatial metrics in the action functional a massless 
scalar (dilaton) field occurs which scales all 
masses including the Planck mass. Instead of the expansion of the 
universe we obtain the Hoyle-Narlikar type of mass evolution, where 
the temperature history of the universe is replaced by the mass 
history. We show that this conformal - invariant cosmological 
model gives a satisfactory description of the new supernova Ia 
data for the effective magnitude - redshift relation without a 
cosmological constant and make a prediction for the high-redshift 
behavior which deviates from that of standard cosmology for 
$z>1.7$. 
\end{abstract} 
\begin{keyword} 
General Relativity and Gravitation, Cosmology, Observational 
Cosmology, Standard Model 
\\[2mm] 
{\sc PACS}: 12.10.-g, 95.30.Sf, 98.80.-k, 98.80.Es 
\end{keyword} 
\end{frontmatter} 
% 
%--------------------start-------------------- 
 
%%%%%%%%%%%%%%%%%%%%%%%% 
\section{Introduction} 
%%%%%%%%%%%%%%%%%%%%%%%% 
% 
The recent data for the luminosity-redshift relation obtained by 
the supernova cosmology project (SCP) \cite{snov} point to an 
accelerated expansion of the universe within the standard 
Friedman-Robertson-Walker (FRW) cosmological model. 
Since the fluctuations of the microwave 
background radiation \cite{10} provide evidence for a flat 
universe a finite value of the cosmological constant $\Lambda$ has 
been introduced \cite{dark} which raises to the cosmic coincidence (or 
fine-tuning) problem ~\cite{9}. 
A most common approach to the solution of this problem is to 
allow a time dependence of the cosmological constant (``Quintessence''
\cite{wetterich,9}), the speed of light \cite{vsl} or the fine 
structure constant \cite{val}. 
 
The present paper is devoted to an alternative description of the 
new cosmological supernova data without a $\Lambda$- term as 
evidence for Weyl's geometry of similarity~\cite{we}, where 
Einstein's theory takes the form of the conformal - invariant 
theory of a massless scalar field~\cite{pct,be,plb1,plb,ps1,pp}. 
 
As it has been shown by Weyl~\cite{we} already in 1918, conformal 
- invariant theories correspond to the relative standard of 
measurement of a conformal - invariant ratio of two intervals, 
given in the geometry of similarity \footnote{ The geometry of 
similarity is characterized by a measure of changing the length of 
a vector on its parallel transport. In the considered dilaton 
case, it is the gradient of the dilaton. In the following, we call 
the scalar conformal - invariant theory the conformal general 
relativity (CGR) to distinguish it from the original 
Weyl~\cite{we} theory where the measure of changing the length of 
a vector on its parallel transport is a vector field (that leads 
to the defect of the physical ambiguity of the arrow of time 
pointed out by Einstein in his comment to Weyl's 
paper~\cite{we}).} as a manifold of Riemannian geometries 
connected by conformal transformations. This ratio depends on nine 
components of the metrics whereas the tenth component became the 
scalar dilaton field that can not be removed by the choice of the 
gauge. In the current literature~\cite{pr,kl} (where the dilaton 
action is the basis of some speculations on the unification of 
Einstein's gravity with the standard model of electroweak and 
strong interactions including modern  theories of supergravity) 
this peculiarity of the conformal - invariant version of 
Einstein's dynamics has been overlooked. 
 
The energy constraint converts this dilaton into a time-like classical 
evolution parameter which scales all masses including the Planck mass. 
In the conformal cosmology (CC), the evolution of the value of the massless 
dilaton field (in the homogeneous approximation) corresponds to that of the 
scale factor in standard cosmology (SC). 
Thus, the CC is a field version of the Hoyle-Narlikar 
cosmology \cite{NR}, where the redshift reflects the change of the atomic 
energy levels in the evolution process of the elementary particle masses 
determined by that of the scalar dilaton field \cite{plb,NR,bbppz}. 
The CC describes the evolution in the conformal time, which 
has a dynamics different from that of the standard Friedmann model. 
 
In the present paper we will discuss as an observational 
argument in favour of the CC scenario that the Hubble diagram 
(effective magnitude - redshift- relation: $m(z)$) including the recent SCP 
data \cite{snov} can be described without a cosmological constant.

%%%%%%%%%%%%%%%%%%%%%%%%%%%%%%%%%%%%%% 
\section{Conformal General Relativity} 
%%%%%%%%%%%%%%%%%%%%%%%%%%%%%%%%%%%%%% 
 
The principle of relativity of all standards of measurement 
%%~\cite{grg}-\cite{114}. 
can be incorporated into the unified theory 
through the Weyl geometry of similarity as 
a manifold of conformal - equivalent Riemannian geometries. 
To escape defects of the first Weyl version of 1918~\cite{we}, 
we use the scalar-tensor conformal invariant 
$(\hat g_{\mu\nu}= w^2 g_{\mu\nu})$ where $w$ is a 
dilaton scalar field described by the 
Penrose-Chernikov-Tagirov (PCT) action~\cite{pct} 
\begin{eqnarray} 
S_{CGR}&=&-\int\limits_{ }^{ }d^4x\sqrt{-\hat g}\frac{1}{6}R(\hat g) 
\nonumber\\ 
&=&\int d^4x\left[-\sqrt{-g}\frac{w^2}{6} R(g)+ 
w \partial_{\mu}(\sqrt{-g}g^{\mu\nu}\partial_{\nu}w )\right]~ 
\end{eqnarray} 
with negative sign. The action and conformal - invariant equations 
of this theory coincide with the ones of Einstein's general 
relativity (GR) expressed in terms of the conformal - invariant 
Lichnerowicz variables $F_{(n)}$, including the metric 
$g$~\cite{L} \be \label{L} 
F^L_{(n)}=||{}^{(3)}g||^{-n/6}F_{(n)}~,~~~~~ (ds^L)^2= 
g^L_{\mu\nu}dx^{\mu}dx^{\nu},~~~~||{}^{(3)}g^L||=1~, \ee where 
${}^{(3)}g_{ij}$ are the 3-dimensional metric components, $(n)$ is 
the conformal weight for a tensor $(n=2)$,~vector $(n=0)$, spinor 
$(n=-3/2)$,~and scalar $(n=-1)$ field. The role of the dilaton field in 
GR is played by the scale-metric field \be\label{scale} 
w_g=||{}^{(3)}g||^{1/6} M_{\rm Planck}\sqrt{\frac{3}{8\pi}}~. \ee 
Therefore, we call this theory the conformal general relativity 
(CGR). 
 
In contrast to Einstein's general relativity theory, in Weyl's conformal 
relativity we can measure only 
a ratio of two Einstein intervals that depends only on nine 
components of the metric tensor. This means that the conformal 
invariance allows us to remove only one component of the metric 
tensor using the scale-free Lichnerowicz conformal - invariant field 
variables~(\ref{L}). 
We show that the conformal invariance of the 
action, the variables, and the measurable quantities gives us an 
opportunity to solve the problems of modern cosmology without 
inflation by the definition of the observables  as conformal - invariant 
quantities. 
We introduce the conformal time, the conformal (coordinate) 
distance, the conformal density, the conformal pressure, etc. 
using instead of the FRW cosmic scale factor the homogeneous dilaton field 
which scales all masses in the universe. 

After the introduction of 
the CGR for an empty universe we give now to the action of the 
matter fields in a conformal invariant formulation of the 
Standard Model (SM) 
\begin{eqnarray} 
S_{CSM}&=&\int d^4x \sqrt{-g} 
\left[ 
\frac{|\Phi|^2}{6}R(g) 
+ {\cal L}^{SM}_{0}(g, \{v_i\}, \{\psi_j\}, \Phi) 
+ {\cal L}_{\rm Higgs}(|\Phi|, w)\right]~, 
\end{eqnarray} 
where ${\cal L}^{SM}_{\lambda}(g, \{v_i\}, \{\psi_j\}, \Phi)$ 
is the  SM Lagrangian with the metric tensor $g$, the 
Higgs field $\Phi$,
the vector boson fields $\{v_i\}$, the spinor fields $\{\psi_j\}$ and 
the coupling constant $\lambda$ of the conventional Higgs potential. 
The latter one has to be replaced by the conformal - invariant one 
\be 
{\cal L}_{\rm Higgs}(\Phi, w)= 
-\lambda\left[ \left(|\Phi|\right)^{2} - C^2(w) \right]^{2}~, 
\ee 
where the mass term of the Higgs field $C(w) = y_{\rm Higgs}w$ 
is rescaled by the cosmological dilaton $w$. The 
conformal - invariant interactions of the dilaton and the Higgs 
doublet form the effective Newton coupling in the gravitational 
Lagrangian \be \frac{|\Phi|^2-w^2}{6}R~. \ee From this term the 
necessity becomes obvious to introduce the modulus $\phi $ and the 
mixing angle $\chi$ of the the dilaton-Higgs mixing~\cite{vp} as 
new variables by \be 
w=\phi~\cosh\chi,~~~~~~|\Phi|=\phi~\sinh\chi~~, 
%~~~~~~~(|\Phi|^2-W^2=-\phi^2)~, 
\ee 
so that the total Lagrangian of our conformal cosmology model takes 
the form 
\begin{eqnarray}\label{T} 
{\cal L}&=&{\cal L}_{CGR} + {\cal L}_{CSM}\nonumber\\ 
&=& - \frac{\phi^{2}}{6}R -\partial_{\mu}\phi 
\partial^{\mu}\phi + \phi^{2}\partial_{\mu}\chi\partial^{\mu}\chi+ 
{\cal L}_{\rm Higgs}(\phi, \chi)+ \bar \psi_e y_e\phi \sinh\chi \psi_e+... 
~, 
\end{eqnarray} 
where the Higgs Lagrangian 
\be 
{\cal L}_{\rm Higgs}(\phi, \chi)= 
 -\lambda\phi^{4}\left[\sinh^2\chi-y^{2}_{\rm Higgs} 
\cosh^2\chi\right]^{2} 
\ee 
describes the conformal - invariant Higgs 
effect of the spontaneous SU(2)  symmetry breaking 
\be 
\frac{\partial {\cal L}_{\rm Higgs}}{\partial \chi} =0~\Rightarrow 
\chi_1 = 0, ~~~~\sinh \chi_{2,3} = \pm \frac{y_{\rm 
Higgs}}{\sqrt{1 - y^{2}_{\rm Higgs}}}\sim 10^{-17}~ 
\ee 
corresponding to the latter pair of solutions ($\chi_{2,3}$). 
The masses of elementary particles are also 
scaled  by the modulus of the dilaton-Higgs mixing. There are two 
ways to obtain the Standard Model. The simplest way is to use a 
scale transformation to convert this modulus into a constant 
(instead of the Lichnerowicz gauge~(\ref{L})) 
\be 
\label{crgr} 
\phi(x_0,x)=\vh_0 = M_{\rm Planck}\sqrt{\frac{3}{8\pi}}~. 
\ee 
In this case the Lagrangian (\ref{T}) goes over into the 
Einstein-Hilbert one with 
\be 
\label{yh} 
y_{\rm Higgs}=\frac{m_X}{\vh_0}\sim 10^{-17}~. 
\ee 
 
In the limit of infinite Planck mass the SM sector decouples from 
the gravitational one and takes the standard renormalizable form 
with the Higgs potential 
\be 
\label{yh1} 
-\lambda(X^2 - {m}^{2}_X)^{2} + O\left(\frac{1}{M_{\rm Planck}}\right)~, 
\ee 
where the notations 
$\vh_0\chi = X $ and  $\vh_0 y_{\rm Higgs} = {m}_X$ 
have been introduced for the Higgs field and its mass term, respectively. 
However, the gauge (\ref{crgr}) violates the conformal symmetry of the 
equations of motion and introduces an absolute standard of measurement 
of geometric intervals depending on ten components. 
This way leads to the standard cosmology. 
 
The second way is to choose the Weyl relative standard of 
measurement of intervals depending on  nine components of the 
metric tensor in the general case. This way is compatible with 
the Lichnerowicz gauge~(\ref{L}) that does not violate the 
conformal symmetry of the equations of motion in the conformal - 
invariant theory considered. 
In this case, the equality~(\ref{crgr}) 
follows from the energy constraint and means the current 
(non-fundamental) status of Planck mass~\cite{pp}. 
The Weyl relative standard of measurement leads to the conformal 
cosmology~\cite{plb}. 
 
%%%%%%%%%%%%%%%%%%%%%%%%%%%%%%%%%%%%%%%%%%%%%%%%%%%%%%%%%%%%%%%%% 
\section{Cosmological solutions for the dilaton - Higgs dynamics} 
%%%%%%%%%%%%%%%%%%%%%%%%%%%%%%%%%%%%%%%%%%%%%%%%%%%%%%%%%%%%%%%%% 
 
It is well-known that the homogeneous and isotropic approximation to 
GR is described by the metric 
\be
\label{GR} 
ds^2= g_{00}(x_0)dx^0dx^0-a^2(x^0)dx^idx^i= a^2(x^0)(ds^L)^2~,
\ee 
where 
$dt=\sqrt{g_{00}}d x_0$ is the Friedmann time interval. 
 
In this approximation CGR is described by the 
flat conformal space-time 
\be 
\label{cst} 
(ds^L)^2=d\eta^2-dx_i^2~,
\ee
where $d\eta=\sqrt{g^L_{00}}d x_0$ is the conformal time interval
and the abbreviation $N_0=\sqrt{g^L_{00}}$ will be used.  
For simplicity we will restrict us here to the discussion of flat 
space. 
 
The r\^ole of the cosmic scale factor in CGR is played by the 
zero momentum mode of the Fourier decomposition of the dilaton field, 
\be 
\vh(x_0) = \frac{1}{V} \int d^3x \phi^L (x_0,\vec x)~, 
\ee 
that scales (as we have seen before) all masses of elementary particles 
including the Planck mass. 
The infrared interaction of the complete set of local 
independent variables $\{f\}$ with this dilaton zero mode 
$\vh(x_0)$ is taken into account exactly, and it is the subject of 
the well-known problem of the cosmological creation of particles 
in terms of the conformal variables~(\ref{L}), see also~\cite{pzg}. 
From the CGR action, we obtain the equation of motion for the dilaton 
field as the conformal analogue of the Friedmann equation for the 
evolution of the universe 
\be
\label{cce} 
\frac{\partial S}{\partial N_0} = 0~~~~~ 
\Rightarrow~~~~~ (\vh')^2 = \rho(\vh)~, 
\ee 
where the prime denotes the derivative with respect to the conformal time
$\eta$. 
\be 
\rho(\vh) = \frac{1}{V}\int d^3x T_{00}(\vh)
\ee 
is the conformal energy density which is connected with the SC one by
$\rho(\vh)=a^4~\rho_{SC}~$, where $a=\vh/\vh_0$. 
 
The cosmic evolution of dilaton masses leads to the redshift of 
energy levels of star atoms~\cite{NR} as a function of the elapsed 
conformal time. 
We can introduce the Hubble parameter of the CC model,
$H_0=\vh'(\eta_0)/\vh(\eta_0)$, 
which can be used to fix the integration constant occuring in the solution 
of the evolution equation (\ref{cce}) with the present day value 
of the dilaton $\vh_0=\vh(\eta_0)$
\be 
\vh_0= \frac{\sqrt{\rho(\vh_0)}}{H_0}= M_{\rm 
Planck} \sqrt{\frac{3}{8\pi}}~. 
\ee 
In CC, the Planck mass is subject to cosmic evolution
and thus not a fundamental parameter that could be used to describe
the beginning of the Universe. 
 
The field theory reproduces all regimes of the classical SC in 
their conformal versions. In particular, the theory of the free 
field describes all the equations of state that are known in the 
standard cosmology: the rigid state 
($p_{\rm Rigid} = \rho_{\rm Rigid}(\vh)= {\rm const}/\vh^2$), 
the radiation state 
($p_{\rm Radiation} = \rho_{\rm Radiation} / 3 = {\rm const}$), 
and the matter state 
($p_{\rm Matter} = 0, ~~\rho_{\rm Matter} = {\rm const} \vh $)~\cite{ps1,pp}. 
The origin of the rigid state are excitations of 
the homogeneous graviton and the dilaton - Higgs field mixing; 
the radiation state corresponds to excitations of other massless fields and 
the matter one to those of massive fields. 
 
Now we can ask: What is the best regime for a description of 
the latest Supernova data on the luminosity distance - redshift relation 
and is this regime compatible with the other cosmological data, like
the CMB radiation and element abundances?
 
%%%%%%%%%%%%%%%%%%%%%%%%%%%%%%%%%%%%%% 
\section{Luminosity distance - redshift relation} 
%%%%%%%%%%%%%%%%%%%%%%%%%%%%%%%%%%%%%% 
 
Let us establish the correspondence between the SC and the CC 
determined by the evolution of the dilaton~(\ref{cce}), where the 
time $\eta$, the density $ \rho(\vh)$, and the Hubble parameter 
$H_0$ are treated as measurable quantities. 
Let us introduce the standard cosmological definition 
of the redshift and density parameter 
\bea 
\label{denspar} 
1+z \equiv \frac{1}{a(\eta)}=\frac{\vh_0}{\vh(\eta)}~,%\\ \nonumber 
~~~\Omega(z)=\frac{\rho(\vh)}{\rho(\vh_0)} , 
\eea 
where $\Omega(0)=1$ is assumed. The density parameter $\Omega(z)$ 
is determined  in both the SC and the CC as 
\be 
\label{coc} 
\Omega(z)=\Omega_{\rm Rigid}(1+z)^2+\Omega_{\rm 
Radiation}+ \frac{\Omega_{\rm Matter}}{(1+z)}
+ \frac{\Omega_{\Lambda}}{(1+z)^4}~. 
\ee 
We added here the $\Omega_{\Lambda}$-term that corresponds to the 
$\lambda\vh^4$ interaction in the conformal action in order to have the 
complete analogy with the standard cosmology. 
Then the equation~(\ref{cce}) takes the form 
\be
\label{etad} 
H_0\frac{d\eta}{dz}=\frac{1}{(1+z)^2}\frac{1}{\sqrt{\Omega(z)}}~, 
\ee 
and determines the dependence of the conformal 
time on the redshift factor. This equation is valid also 
for the conformal time - redshift relation in the SC 
where this conformal time is used for description of a light ray. 
 
A light ray traces a null geodesic, i.e. a path for which the 
conformal interval $(ds^L)^2=0$ thus satisfying the equation 
${dr}/{d\eta} = 1$. As a result we obtain for the coordinate 
distance as a function of the redshift 
\be
\label{rdi} 
H_0 
r(z)=\int_0^z \frac{dz'}{(1+z')^2}\frac{1}{\sqrt{\Omega(z')}}. 
\ee 
The equation~(\ref{rdi}) coincides with the similar relation 
between coordinate distance and redshift in SC. 
 
In the comparison with the stationary space in SC and stationary 
masses in CC, a part of photons is lost. To restore the full 
luminosity in both SC and CC we should multiply the coordinate 
distance by the factor $(1+z)^2$. This factor comes from the 
evolution of the angular size of the light cone of emitted photons 
in SC, and from the increase of the angular size of the light cone 
of absorbed photons in CC. 
 
However, in  SC  we have an additional factor $(1+z)^{-1}$ due to 
the expansion 
of the universe, as measurable distances in SC are related to 
measurable distances in CC (that coincide with the coordinate ones) 
by the relation 
\be 
\ell={a}\int\limits_{ }^{ }\frac{dt}{a}=\frac{r}{1+z}~.
%,~~~a=\frac{\vh}{\vh_0}. 
\ee 
Thus we obtain the relations 
\be \label{scr} 
\ell_{\rm SC}(z) = (1+z)^2 \ell = (1+z)  r(z)~, 
\ee 
\be \label{ccr} 
\ell_{\rm CC}(z) =  (1+z)^2  r(z)~. 
\ee 
This means that the observational data are described by different 
regimes in SC and CC. For example, the rigid state 
(i.e. $\Omega_{\rm Rigid}=1$) gives the relation 
\be \label{ccrr} 
\ell_{\rm CC}(z) = z+\frac{z^2}{2}~. 
\ee 
 
\begin{figure}[ht] 
\epsfig{figure=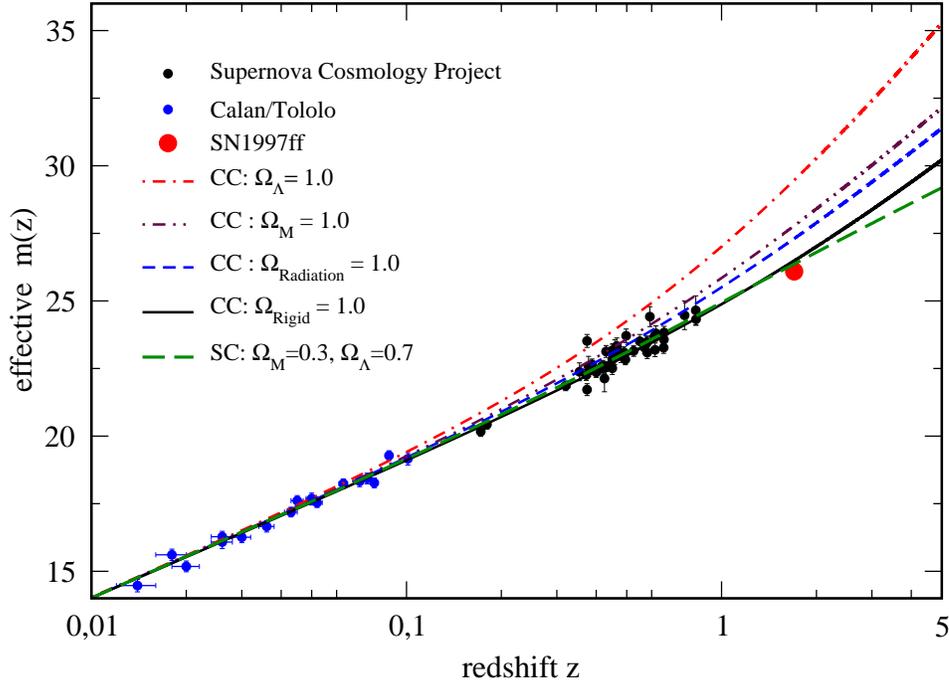,width=16cm,angle=0} 
\caption{$m(z)$-relation for a flat universe model in SC and CC. 
The data points include those from 42 high-redshift Type Ia 
supernovae~\protect\cite{snov} and that of the recently reported 
farthest supernova SN1997ff~\protect\cite{sn1997ff}. An optimal 
fit to these data within the SC requires a cosmological constant 
$\Omega_{\Lambda}=0.7$, whereas in the CC 
 these data require the dominance of the rigid state.
\label{fig1}} 
\end{figure}

In Fig.~\ref{fig1} we compare the results of the SC and CC for the 
effective magnitude-redshift relation: 
$m(z) = 5 \log{[H_0\ell(z)]} + {\cal M}$~, where ${\cal M}$ is a 
constant, with recent experimental data for distant supernovae
\cite{snov,sn1997ff}. Within the CC model the pure rigid state of 
dilaton-Higgs dynamics without cosmological constant gives the best 
description  and is equivalent to the SC fit up to the SN1997ff 
point.   
 
%%%%%%%%%%%%%%%%%%%%%%%%%%%%%%%% 
\section{Cold Universe Scenario} 
%%%%%%%%%%%%%%%%%%%%%%%%%%%%%%%% 

In this section we want to discuss the consistency of the here
described CC scenario of a nonexpanding Universe, in which the 
observed redshift of spectra is due to time-dependent elementary 
particle masses, with other cosmological observations such as
the CMB radiation and the distribution of elements. 

In the limit of the Early Universe, $\vh \Rightarrow 0$, 
the CGR action also gives the most singular 
rigid state 
${\rho}/{\rho_0}=\Omega_{\rm Rigid}(z+1)^2$ 
and the primordial motion of the dilaton described before 
\bea 
\vh^2(\eta)&=&\vh^2_I[1+ 2 H_I \eta]= 
\frac{\vh^2_0}{(1+z)^2}~,\nonumber \\ 
H(z)&=&\frac{\vh'}{\vh}=H_0(1+z)^2~. 
\eea 
At the point of coincidence of the Hubble parameter of this motion 
with the mass of vector bosons $m_v(z)\sim H(z)$, there occurs the 
intensive creation of longitudinal vector bosons, see \cite{b+01}. 
Fast thermal equilibration of this boson system takes place since
for the inverse relaxation time holds
$\eta^{-1}_{\rm relaxation}=\sigma_{\rm scat.}n_v\geq 
H(z)$, and therefore the density of created vector bosons $n_v$ 
defines an equilibrium temperature which appears to be an the integral 
of motion of the cosmic evolution 
$ T_{\rm eq}\!\simeq\! [m^2_v(z)H(z)]^{1/3}\!\simeq\! (m_W^2H_0)^{1/3}
=2.7 K \!\sim\! H_I $. 
This is a surprisingly good agreement of $T_{\rm eq}$ with the CMB radiation 
temperature. 
 
It is worth to emphasize this difference between the CC model and the SC 
ones: In conformal cosmology, the CMB temperature remains constant (cold 
scenario) but the masses evolve throughout the history of the universe 
due to the time dependence of the dilaton field 
\be 
\label{mz} 
m_{\rm era}{(z_{\rm era})}=\frac{m_{\rm era}(0)}{(1+z_{\rm era})}
=T_{\rm eq}~, 
\ee 
where $m_{\rm era}(0) $ is the present-day value of a characteristic 
energy (mass) scale determining the onset of an 
era of the universe evolution. 
 
Eq. (\ref{mz}) has the important consequence that all those 
physical processes which concern the chemical composition of the 
universe and which depend basically on Boltzmann factors with the 
argument $(m/T)$ cannot distinguish between the mass history of conformal 
cosmology and the temperature history of standard cosmology due to the 
relations 
\be\nonumber 
\frac{m(z)}{T(0)}=\frac{m(0)}{(1+z)T(0)}=\frac{m(0)}{T(z)}~. \ee 
This formula makes transparent that in this order of approximation 
a $z$-history of masses with invariant temperatures in the rigid 
state of CC is equivalent to a $z$-history of 
temperatures with invariant masses in the radiation stage of SC. 
We expect therefore that the conformal 
cosmology will be as successful as the 
standard cosmology in the radiation stage for describing, e.g., the 
neutron-proton ratio and the primordial element abundances. 
 
An important new feature of the conformal cosmology relative to 
the standard one is the absence of the Planck era, since the 
Planck mass is not a fundamental parameter but only the 
present-day value of the  dilaton field~\cite{plb}. 
 
%%%%%%%%%%%%%%%%%%%% 
\section{Conclusion} 
%%%%%%%%%%%%%%%%%%%% 
 
We have presented an approach according to which the new supernova 
data can be interpreted as evidence for a new type of geometry 
in Einstein's theory rather then a new type of matter. This 
geometry corresponds to the relative standard of measurement and 
to a conformal cosmology with constant three-volume. 
In this cosmology, the dilaton field scales all masses and 
its evolution is responsible for observable phenomena like the redshift 
of spectra from distant galaxies.
The evolution of all masses replaces the familiar evolution of the scale 
factor in standard cosmologies.
The infrared dilaton - elementary particle 
interaction leads to particle creation 
\cite{b+01} and in turn to the occurence of the CMB radiation with a
temperature of $2.7$ K not changed ever since. 

We have defined the cosmological parameters in the 
conformal cosmology, and we have found that the effective 
magnitude - redshift relation (Hubble diagram) for a rigid state 
which originates from the dilaton - Higgs dynamics
describes the recent observational data for distant 
(high-redshift) supernovae including the farthest one at $z=1.7$. 
While in the standard FRW 
cosmology interpretation a $\Lambda$ - term (or a quintessential 
analogue) is needed, 
which entails a transition from decelerated to accelerated 
expansion at about $z\sim 1.7$, the cosmology presented here does not need
a $\Lambda$ - term. 
Both cosmologies make different predictions for the 
behaviour at $z > 1.7$. 
Provided that the CSM with a Higgs potential
gives a correct description of the matter sector, our findings suggest 
that new data at 
higher redshift could discriminate between the alternative 
cosmological interpretations of the luminosity - redshift relation 
and answer the question: Is the universe expanding or not? 
\subsection*{Acknowledgement} 
We thank Dr. A. Gusev and Prof. S. Vinitsky for fruitful discussions. 
One of us (V.N.P.) acknowledges support by the Ministery for Education, 
Science and Culture in Mecklenburg - Western Pommerania and thanks for the 
hospitality of the University of Rostock where this work has been completed. 
D.P. thanks the RFBR (grant 00-02-81023 Bel$\_a$) for support. 
 
%______________________________ References ______________________________ 

\end{document}